# DAEMON: Dataset/Platform-Agnostic Explainable Malware Classification Using Multi-Stage Feature Mining

RON KORINE AND DANNY HENDLER, (Member, IEEE)
Department of Computer Science, Ben-Gurion University of the Negev, Be'er Sheva 84105, Israel

Corresponding author: Ron Korine (ronkor@post.bgu.ac.il)

This work was supported in part by the Cyber Security Research Center, Ben-Gurion University of the Negev.

**ABSTRACT** Numerous metamorphic and polymorphic malicious variants are generated automatically on a daily basis. In order to do that, malware vendors employ mutation engines that transform the code of a malicious program while retaining its functionality, aiming to evade signature-based detection. These automatic processes have greatly increased the number of malware variants, deeming their fully-manual analysis impossible. Malware classification is the task of determining to which family a new malicious variant belongs. Variants of the same malware family show similar behavioral patterns. Thus, classifying newly discovered malicious programs and applications helps assess the risks they pose. Moreover, malware classification facilitates determining which of the newly discovered variants should undergo manual analysis by a security expert, in order to determine whether they belong to a new family (e.g., one whose members exploit a zero-day vulnerability) or are simply the result of a concept drift within a known malicious family. This motivated intense research in recent years on devising high-accuracy automatic tools for malware classification. In this work, we present DAEMON—a novel dataset-agnostic malware classifier. A key property of DAEMON is that the type of features it uses and the manner in which they are mined facilitate understanding the distinctive behavior of malware families, making its classification decisions *explainable*. We've optimized DAEMON using a large-scale dataset of x86 binaries, belonging to a mix of several malware families targeting computers running Windows. We then re-trained it and applied it, without any algorithmic change, feature re-engineering or parameter tuning, to two other large-scale datasets of malicious Android applications consisting of numerous malware families. DAEMON obtained highly accurate classification results on all datasets, establishing that it is not only dataset-agnostic but also *platform-agnostic*. We analyze DAEMON's classification models and provide numerous examples demonstrating how the features it uses facilitate explainability.

## I. INTRODUCTION

Traditional anti-malware software relies on signatures to uniquely identify malicious files. Signatures of files are based on their content. Malware vendors have responded by developing metamorphic and polymorphic malware. These malware are generated automatically using mutation engines that apply one or more obfuscation techniques. Common obfuscation techniques include subroutine reordering, dead-code insertion, register renaming, and encryption.

The associate editor coordinating the review of this manuscript and approving it for publication was Juan Liu.

These techniques transform the code of a malicious program, while retaining its functionality, in order to evade signature-based detection [1]–[4].

These mechanisms for automatic malware generation caused the number of malware variants to skyrocket. The number of new variants created during 2016-2018 alone is estimated by more than 1.25 billion [5], deeming manual analysis of new variants infeasible. Variants belong to the same *malware family* if they show similar behavior and attempt to exploit the same vulnerabilities. This often implies that they are metamorphic/polymorphic variants of the same original malicious program. *Malware classification* is the task of determining to which family a new variant belongs. Automatic classification of newly discovered variants helps assess the risks they pose and determine which ones should undergo manual analysis. Security expert's analysis helps determine whether they belong to a new family or are malware variants indicating a concept drift within a known malicious family. The process of manual analysis is time-consuming and costly. Thus, malware classification is highly beneficial because it can greatly reduce the number of samples that require manual analysis. Consequently, high-accuracy automatic tools for malware classification are a key component in cyber security.

Malware classification is based on features extracted from analyzed malware samples. Tremendous efforts have been invested in recent years in classifying malware based on features extracted using *static analysis* (e.g. [6]–[17]), *dynamic analysis* (e.g. [18]–[24]), or hybrid techniques that utilize both static and dynamic features (e.g. [25]–[28]). Whereas static analysis is based solely on the contents of the sample under consideration, dynamic analysis executes it in a controlled environment and studies its run-time behavior. Although dynamic analysis does not require disassembly of the executable sample, it consumes much more time and computing resources in comparison with static analysis [29]. Moreover, malware vendors have found ways of hindering, impeding, and evading dynamic analysis [1], [2]. In this work, we focus on malware classification using static analysis.

We say that a classifier is *dataset-agnostic* if we can apply it to different datasets without performing any algorithmic changes, feature re-engineering, or parameter tuning. We say that a feature derived from an analyzed sample is *platform-agnostic*, if it does not rely on any knowledge of the platform which the sample targets. This implies that the computation of a platform-agnostic feature must be done without any knowledge of the sample's executable-format or the platform's instruction set architecture. If a feature is not platform-agnostic, we say it is *platform-dependent*. When we refer to the lack of feature re-engineering, we mean that features are selected according to DAEMON's feature mining algorithm which works in exactly the same manner when applied to different datasets. *Obviously, in general, different features would be selected for different datasets*.

Examples of platform-dependent features include the distribution of instruction opcodes, platform register usage frequency, strings that appear in a specific header of the executable, the number of executable sections and their sizes, etc. Examples of platform-agnostic features include $N$-grams, sample size, and features derived from the distribution of byte-values and from the entropy of the sample's contents. We call a malware classifier *platform-agnostic* if it can accurately classify collections of malware executables, regardless of the platform they target, without performing any algorithmic changes or any form of feature re-engineering. Consequently, platform-agnostic classifiers are classifiers that only use platform-agnostic features.

## II. CONTRIBUTIONS

To the best of our knowledge, no previous effective and explainable malware classifier was evaluated on several datasets of executables targeting different computer platforms, let alone, without performing some algorithmic changes, feature re-engineering[1] or parameter tuning.

In this work, we present DAEMON, the first provably effective and explainable platform-agnostic and dataset-agnostic malware classifier. We have optimized DAEMON using Microsoft's Kaggle Malware Classification Challenge dataset [30], which consists of 21,741 malware samples of Portable Executable (PE) format. The dataset is comprised of x86 binaries,[2] belonging to a mix of 9 different families. DAEMON provides classification results which place it among the top 3 out of more than 370 classifiers evaluated on this dataset. We then re-trained and applied DAEMON, *without any algorithmic change, feature re-engineering or parameter tuning*, to two other datasets that are collections of Dalvik bytecode Android applications. The first is the Drebin dataset [17], [31], consisting of 5,560 malicious applications from 179 different malware families. The second is the CIC-InvesAndMal2019 dataset [32], [33], which consists of 426 malicious files belonging to 42 malicious families. DAEMON's classification results significantly exceed those of all previous classifiers evaluated on both Drebin and CIC-InvesAndMal2019.

DAEMON's high classification accuracy of executables from different platforms stems from the fact that it considers *all $N$-grams* for certain values of $N$ that are much larger than those typically used by malware classifiers as potential features. Since for large values of $N$ the set of all $N$-grams that appear in dataset files is huge, the key challenge addressed by DAEMON is that of efficiently mining a relatively small subset of effective features from this set. As we describe in section IV, these features are mined from the initial set of candidate features in several stages. Specifically, feature mining is done in a manner that attempts to preserve a sufficient number of high-quality separating features, *for every pair of malware families*.

An important advantage which arises from the type of features used by DAEMON and the manner in which they are mined is that its classification results are *explainable*. Each feature that "survives" the filtering process and is used by the machine learning algorithm employed by DAEMON is labeled by the set of family-pairs that it is effective in separating. Moreover, the long $N$-grams used by DAEMON as features are often either readable strings (e.g., strings identifying imported API functions) or snippets of malicious code. The combination of information-rich features and the knowledge of which families they are able to tell apart facilitates the analysis of a malware family's behavior and what distinguishes it from other families. This is in contrast with many malware classifiers, which are based on statistical features. DAEMON's code is publicly available.[3]

---

[1] As we've written previously, this *does not* mean that the same features are used for different datasets.

[2] This is the format used by Windows executables.

The rest of this article is organized as follows. We describe related work in Section III. We then present the DAEMON classifier in Section IV. In Section VI, we describe the datasets we use in this work. This is followed by a description of our experimental evaluation and its results in Section VII. We provide examples of how DAEMON's classification results allow gaining insights into the behavior of malware families in Section IX. We discuss more related work, limitations and avenues for future work in Section X, and conclude in Section XI.

## III. RELATED WORK

Malware classification is one of the key challenges in contemporary cyber security research. This task can be addressed using one of the following approaches: *static analysis*, *dynamic analysis*, or *hybrid analysis* which is a combination of the two. Static analysis is based solely on the analysis of an executable file's contents, whereas dynamic analysis is based on the run-time behavior of a program as it executes. Hybrid analysis is not very common for classification purposes and is used in a relatively small number of works [26]–[28], [34], [35]. In the following, we describe several prior works that presented static or dynamic analysis malware classifiers and focus on those works that were evaluated on the three datasets on which we evaluated DAEMON.

### A. DYNAMIC ANALYSIS

Most contemporary malware employs obfuscation techniques such as encryption and packing in order to make static analysis difficult [2], [3]. Consequently, many works take the approach of developing behavior-based malware detection and classification methods. Dynamic analysis is based on the run-time behavior of the malware, typically executed inside a secure sandbox. Therefore, it is unaffected by such obfuscation methods. On the downside, dynamic analysis consumes more computational resources [29]. Moreover, contemporary malware often checks whether it is running in a virtual environment and exposes its malicious nature only after verifying that this is not the case. Consequently, it becomes increasingly difficult to devise virtual environments that seem sufficiently genuine for the malware to expose its payload [1].

Techniques for evading dynamic analysis have existed for many years and are utilized not only by malicious programs that target Windows-based platforms but also by Android malware. For example, the Android.Adrd Trojan [36], discovered in 2011, executes itself only if either:

- Twelve hours passed since the Android OS was booted.
- The device lost and then re-gained network connectivity.
- A phone-call was received.

In order to aid Android malware researchers, several datasets consisting of samples belonging to various malware families were published. A few of these datasets, such as the Drebin dataset [17] and the Android Malware Genome Project [37], became benchmarks for the evaluation of Android malware classifiers and detectors.

Afonso *et al.* [18] proposed a set of statistical features regarding the behavior of Android applications. These features include frequencies of API-calls and system-calls. In addition, they compared different machine learning models using these features and established that their random forest model substantially outperformed the rest.

Dash *et al.* [22] presented DroidScribe in 2016, a classifier for Android malware that is based on behavioral aspects. These include features such as API-calls and ''high level behaviors'', representing combinations of traditional OS operations (such as process creation) and selected Android methods (such as sending SMS messages). They evaluated their work on the Drebin dataset [17] and the Android Malware Genome Project [37] (which was incorporated into Drebin since then).

Additional works that have taken a dynamic analysis approach for malware classification and used Drebin include [19], [23], [24].

Martin *et al.* [23] combined dynamic analysis and Markov-chain modeling for malware classification. In this work, both classical machine learning classifiers and deep-learning classifiers were used. Massarelli *et al.* [24] analyzed malicious Drebin instances and computed classification features based on an application's resource consumption over time.

Cai *et al.* [19] observed that most dynamic approaches rely on characterization of system calls which are subject to system-call obfuscation. Therefore, their proposed solution, DroidCat, relies on a set of dynamic features. These features include method calls and inter-component communication (ICC), even those defined by user code, third-party libraries, etc., instead of monitoring system calls. They state that their solution has superior robustness to obfuscations in comparison with most state of the art Android malware detection tools (both static and dynamic) at the time.

Several works have applied dynamic analysis techniques for malware classification of PE files. Huang *et al.* [38] performed dynamic analysis for feature extraction and then used a deep-learning classifier for malware family classification. They trained and tested their classifier using a large dataset comprising 6.5 million files. Tian *et al.* [39] extracted API call features via dynamic analysis, devising both a malware detector and a malware family classifier.

They evaluated their algorithm using a dataset comprised of 1,368 malicious and 456 clean PE files. We are not aware of any dynamic analysis work that was done using Microsoft's Kaggle dataset because, as we've mentioned, Microsoft removed the headers from the PE executables in this dataset.

### B. STATIC ANALYSIS
In general, static analysis consumes less computational resources than dynamic analysis, is commonly used, and is able to provide good classification results even when malicious programs are obfuscated. In what follows, we focus on recent works evaluated on common malware datasets including Microsoft's dataset and the Drebin dataset. We start by describing some of the malware classifiers that were evaluated using Microsoft's Kaggle Malware Classification Challenge dataset.[4] To the best of our knowledge, all of these classifiers were evaluated only on this single dataset.

---
[3]https://github.com/RonsGit/DAEMON-Extraction-Process

The winning team in the competition was able to train a powerful model with 99.83% accuracy over 4-folds cross validation, and an extremely small logloss of 0.00283 on the competition's test data [40]. Many works that were published after the competition has ended tested various static-analysis based malware classifiers using this dataset and we proceed to briefly describe a few of them.

Ahmadi *et al.* [11] designed a classifier that combines several types of features derived from pixel intensity, op-code counts, PE metadata, 1-grams, and more. Zhang *et al.* [13] devised an ensemble classification model, obtaining a logloss of only 0.00426 and high cross-validation accuracy of 99.79. Another interesting technique for improving classification accuracy is seen in the work of Hu *et al.* [12], that incorporated threat intelligence data such as anti-virus labels of the malware.

We next describe Android malware detectors and classifiers. Avdieenko *et al.* [8] proposed using the level of dissimilarity between malicious and benign applications in order to detect malware variants. They developed MUDFLOW, a tool that uses sensitive data flows as features to describe the behavior of an application, improving malware detection quality in comparison with earlier works. They evaluated their tool on malware files from the Android Malware Genome Project (which was incorporated into Drebin since then).

Conti *et al.* [6] presented MaMaDroid, a malware detector that attempts to statically profile an application's behavior, in the form of a Markov chain, from the sequence of abstracted API-calls performed by it. Subsequently, it uses this behavioral model to extract features and perform detection. By abstracting API-calls to their packages or families, MaMaDroid gains some resilience to API changes. They evaluated their work on malware files originating from Drebin and VirusShare.[5]

Suarez-Tangil *et al.* [7] raised the problem of obfuscations, and suggested using resource-centric features in order to improve their detector's ability to mitigate obfuscated malware. The resource-centric features are diverse, and include certificates, shared libraries, etc. The authors state that a combination of resource-centric features along with common syntactic features such as strings, services, and APIs, improves their solution's resilience to obfuscations while retaining fine performance. They evaluated their malware detector and family classifier, DroidSieve, on several datasets including the Drebin dataset.

Only a few works devised classifiers that use only platform-agnostic features. Kebede *et al.* [14], Narayanan *et al.* [16] and Le *et al.* [15] all presented deep learning classifiers based on a feature set including mostly N-grams and evaluated them using Microsoft's dataset. DAEMON obtains better results in terms of both logloss and accuracy in comparison with these works.

Although the vast majority of previous works did not use long $N$-gram features for malware detection/classification, as done by DAEMON, a few exceptions exist. Dinh *et al.* [41] employed the Smith-Waterman DNA sequence alignment algorithm in order to generate family signatures. They have found a few interesting sequences in the `Ramnit` and `Lollipop` families of Microsoft's dataset. However, they stated that their algorithm took an extremely long time to run and was thus unable to process all family variants, hence did not manage to fully construct family signatures. DAEMON managed to extract parts of the sequences found by their algorithm efficiently and used them as features for family classification.

Another work that employs long N-grams (128-bytes long) is that of Faruki *et al.* [42], which presents the AndroSimilar malware detector. AndroSimilar targets the detection of zero-day Android malware, using what they call "statistically improbable features". These are long N-grams, whose random occurrence is very improbable. They use these features for generating file signatures. These signatures are then used in order to search for similar (up to a certain threshold) signatures in a designated malware database. The system alerts when a match is found. The most popular 128 features found are stored in a Bloom filter representing the file signature. DAEMON also uses long N-grams, but they are used for family classification rather than for detection. Moreover, whereas AndroSimilar uses a uniform N-gram length, settles for partial matches, and uses fuzzy hashing, DAEMON uses several $N$-gram lengths and requires exact matches.

## IV. THE DAEMON MALWARE CLASSIFIER

The high-level structure of DAEMON's model generation process is presented in Figure 1. The number displayed below each outgoing edge at the bottom of Figure 1 is the number of candidate features that remain after the corresponding algorithm stage, when DAEMON is trained on Microsoft's Kaggle dataset (see Section VI for a description of this dataset). DAEMON's high-level pseudo-code is presented in Algorithm 1.

The key features used by DAEMON are byte-sequence $N$-grams, which are contiguous sequences of $N$ bytes from a sample's content. Using a parameter $C_{max}$, DAEMON builds a set $\mathcal{L} = \{2^i | i \in \{2, 3 \ldots C_{max}\}\}$ of lengths for $N$-grams that will be mined by the algorithm. We would like $C_{max}$ to be large enough to capture large snippets of code, strings, etc., but not too large, since DAEMON's memory consumption and run-time requirements quickly grow with $C_{max}$. We chose to use $C_{max} = 5$, hence DAEMON mines $N$-grams for $N \in \mathcal{L} = \{4, 8, 16, 32\}$. This value of $C_{max}$ was empirically shown to provide good classification results on Microsoft's Kaggle dataset, while requiring reasonable computational resources.

---

[4]https://www.kaggle.com/c/malware-classification/leaderboard
[5]https://virusshare.com

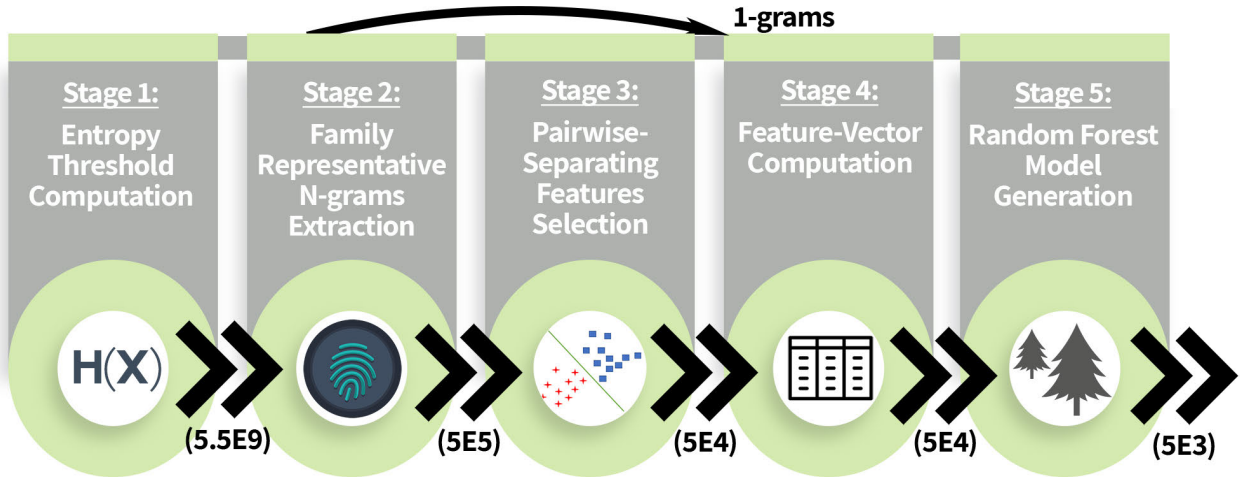

**FIGURE 1.** DAEMON's model generation process.

**Algorithm 1** DAEMON Model Generation Pseudo-Code
1: // Stage 1: Entropy threshold computation.
2: **for** $N \in \mathcal{L}$ **do**
3:    Randomly sample $\alpha \cdot |TrainSet|$ training-set files
4:    Randomly select $\beta$ $N$-grams from each sampled file
5:    $AvgEnt \leftarrow$ average entropy of selected $N$-grams
6:    $t_N \leftarrow AvgEnt \cdot fact[N]$
7: **end for**
8: // Stage 2: Family representative N-grams extraction.
9: **for** each malware family $\mathcal{F}$ **do**
10:    **for** each $F \in \mathcal{F}$ in training set **do**
11:      **for** each $N \in \mathcal{L}$ **do**
12:        **for** each N-gram $s \in F$ s.t. $H(s) \geq t_N$ **do**
13:           Increment $Count(F, s)$
14:        **end for**
15:        $FReps(\mathcal{F}) \leftarrow$ all $s$ s.t. $Count(F, s) \geq \lfloor \gamma |\mathcal{F}| \rfloor$
16:      **end for**
17:      Calculate 1-gram vector $< c_0, \ldots, c_{255} >$ for $F$
18:    **end for**
19: **end for**
20: // Stage 3: Pairwise-separating features selection.
21: **for** each families-pair $< F_1, F_2 >$ out of the $k$ families **do**
22:    Select the topmost $B/\binom{k}{2}$ stage-2 separating features
23: **end for**
24: // Stage 4: Feature-vectors computation.
25: **for** each file $f$ in training set **do**
26:    Use Aho-Corasick to find which $n$-grams appear in $f$
27:    Construct $f$'s feature-vector $v$, where $dim(v) = B + 256$
28: **end for**
29: // Stage 5: Random forest model generation.
30: Generate initial random forest model
31: Choose $C \ll B$ most important features
32: Generate final random forest model

The set of all possible $N$-grams of these lengths is huge and its cardinality is $\Omega(256^{32})$. In order to efficiently mine a small subset of effectively-separating features, DAEMON applies a series of stages, each reducing the size of the candidate-features set. These are described in the following.

## V. STAGE 1: ENTROPY THRESHOLD COMPUTATION

DAEMON's feature mining process starts by computing, for each length $N \in \mathcal{L}$, an *entropy threshold*. The entropy of a byte-sequence $N$-gram $S$ is defined as:

$$H(S) = - \sum_{k=0}^{k=255} P(S, k) \cdot \log_2 P(S, k), \qquad (1)$$

where $P(S, k)$ is the fraction of the bytes of $S$ that assume value $k$. An $N$-gram's entropy is a measure of how much information it stores. For example, if all of its bytes assume the same value, then $H(S) = 0$ holds, indicating that $S$ is unlikely to be a useful feature. If each of its bytes assumes a distinct value, then $H(S)$ obtains the maximum entropy value attainable by an $N$-gram. For each length $N \in \mathcal{L}$, we compute a threshold $t_N$ such that all $N$-grams whose entropy is below $t_N$ will be filtered out.

Threshold $t_N$ is computed as follows (see Algorithm 1): A fraction $\alpha$ of training set files are randomly chosen. Then, $\beta$ $N$-grams from random positions are extracted from each such file. The average entropy of the resulting set of $\alpha \cdot \beta$ $N$-grams is computed. Finally, the entropy threshold for $N$-grams is obtained by multiplying the average entropy by a factor $fact[N] > 1$. DAEMON uses $\alpha = 0.1$ and $\beta = 256$. The factors for $N \in \{16, 32\}$ were set to 1.15, whereas the factors for $N \in \{4, 8\}$ were set to 1.05. The rational for setting a larger threshold-factor for larger values of $N$ is that the number of distinct $N$-grams for $N \in \{16, 32\}$ is orders-of-magnitude larger than that of $N \in \{4, 8\}$, hence stricter filtering is required for larger values of $N$.

### A. STAGE 2: FAMILY REPRESENTATIVE N-GRAMS EXTRACTION

The key goal of stage 2 is to extract *representative N-grams* for each malware family $\mathcal{F}$, for each $N \in \mathcal{L}$. An N-gram is a representative for family $\mathcal{F}$, if its entropy passes threshold $t_N$ and if it appears in at least a fraction $\gamma$ of $\mathcal{F}$'s files. DAEMON uses $\gamma = 0.1$. A family's representative $N$-gram appears in a significant portion of its files. Consequently, it is more likely to characterize the family's distinctive behavior than an $N$-gram that only appears in a negligible fraction of the family's files. While scanning the contents of each file, we also compute the number of occurrences for every file 1-gram (line 17). These features will be candidate features, together with $N$-grams for larger values of $N$, in later stages of the algorithm.

### B. STAGE 3: PAIRWISE-SEPARATING FEATURES SELECTION

In stage 3, we further reduce the set of candidate features by selecting a subset $\mathcal{S}$ of size $B$ of the family representatives output by stage 2 (henceforth called *stage-2 N-grams*). DAEMON uses $B = 50,000$ as this empirically gave the best results on Microsoft's Kaggle dataset. Let $k$ be the number of dataset families. $B$ is constructed by greedily selecting, for each of the $\binom{k}{2}$ family-pairs, the top $B/\binom{k}{2}$ $n$-grams for separating between the two families. The effectiveness of each $n$-gram is measured according to its information gain w.r.t. the pair of families, defined as follows.

Let $F_1, F_2$ be a pair of families. Let $F = F_1 \cup F_2$, $g_1 = |F_1|/|F|$ and $g_2 = |F_2|/|F|$. Then the entropy of $F$ w.r.t. $F_1$, $F_2$ is defined as:

$$H(F, F_1, F_2) = -g_1 \cdot \log_2 g_1 - g_2 \cdot \log_2 g_2. \quad (2)$$

For a stage-2 $n$-gram $s$, let $L(F, s) = \{f \in F | s \in f\}$ and $R(F, s) = \{f \in F | s \notin f\}$. Also, let $gl_1 = |\{f \in L(F, s) | f \in F1\}|/|L(F, s)|$ and $gl_2 = |\{f \in L(F, s) | f \notin F1\}|/|L(F, s)|$. The entropy of $L(F, s)$ w.r.t. $F_1$, $F_2$ is defined as:

$$H(L(F, s), F_1, F_2) = -gl_1 \cdot \log_2 gl_1 - gl_2 \cdot \log_2 gl_2. \quad (3)$$

We define $gr_1$, $gr_2$ using $R(F, s)$ and $H(R(F, s), F_1, F_2)$ similarly. Let $gf_1 = |L(F, s)|/|F|$ and $gr_1 = |R(F, s)|/|F|$. Then the information gain of $s$ w.r.t. $F_1$, $F_2$ is given by:

$$H(F, F_1, F_2) - H(L(F, s), F_1, F_2) \cdot gf_1 \\ - H(R(F, s), F_1, F_2) \cdot gr_1. \quad (4)$$

Note that we do not add the same $n$-gram multiple times, even if it is among the top-most features for multiple pairs. However, we do tag it with all these pairs. A by-product of selecting pairwise-separating features is the following: Each $n$-gram that eventually gets used by DAEMON's detection model is tagged by the set of family-pairs for which it was selected in Stage 3. As we demonstrate in Section IX, this helps in pinpointing the differences between malware families.

### C. STAGE 4: FEATURE-VECTORS COMPUTATION

After stages 1-3 have been completed, $B + 256$ feature-candidates remain: $B$ $N$-grams, for $N \in \mathcal{L} = \{4, 8, 16, 32\}$, as well as 256 features storing the number of occurrences of each 1-gram value for each training set file $f$.

In stage 4, we compute, for each such $f$, a feature-vector of length $B + 256$ to represent $f$. Whereas the latter 256 features were already computed in Stage 2, in Stage 4 we must efficiently find which of the $B$ pairwise-separating $N$-grams are contained in $f$. We do so by applying the Aho-Corasick string-searching algorithm [43], whose complexity is $O\left(|f| + \sum_{i=0}^{i=B-1}(|s_i| + |m_i|)\right)$, where $s_i$ is the length of the $i$'th $n$-gram and $m_i$ is the number of occurrences of $s_i$ found in $f$. We note that the Aho-Corasick algorithm computes the *total* number of occurrences of each $n$-gram in $f$. Nevertheless, the corresponding feature-vector entries are binary: 0 if the corresponding $n$-gram is absent from $f$, or 1, otherwise.[6]

### D. STAGE 5: RANDOM FOREST MODEL GENERATION

Using the feature-vectors output by stage 4, we use Python's Scikit-learn (sklearn) ML library's random forest algorithm for generating an *initial classification model*. We set the number of forest trees to 3,000. With more than 50,000 features, the resulting model is large and tends to overfit. Consequently, we apply to it yet another feature selection stage using *sklearn.feature_selection.SelectFromModel* meta-transformer, for choosing the $C$ initial-model features that have received the highest importance weights. DAEMON uses $C = 5,000$. We then retrain the random forest using the reduced set of features to obtain the *final classification model*.

## VI. DATASETS OVERVIEW

We evaluate DAEMON using three datasets comprised of malware families of two different platforms. The first dataset is Microsoft's Malware Classification challenge dataset [30], consisting of more than 20K Windows Portable Executable (PE) programs. The second is the Drebin dataset [17], consisting of approximately 130K Android Dalvik bytecode executables. The third is CIC-InvesAndMal2019 [32], [33], consisting of 426 malicious Android Dalvik bytecode applications. In the following, we briefly describe each of these datasets.

### A. MICROSOFT's MALWARE CLASSIFICATION CHALLENGE DATASET

This dataset was published in 2015 as part of a Kaggle [44] competition and became a standard benchmark for Windows malware classifiers. It consists of 9 Windows malware families whose names, types, and numbers of training samples are presented in Table 1. In total, there are 10,868 training samples provided along with their class labels. In addition, 10,873 test files are also provided, but without their labels. For each sample, two files are provided: a hexadecimal representation of the sample's PE binary file (without the PE headers) and a disassembly file generated from the binary file using IDA Pro. After training a classification model using the training samples, its performance on the test set is evaluated by uploading to the competition's site a submission file that contains, per every test sample $i$ and class (family) $j$, the probability $p_{i,j}$ that $i$ belongs to $j$ as predicted by the model. In response, the competition's site returns the *multi-class logarithmic loss* (henceforth simply referred to as *logloss*) of the prediction, defined as follows:

$$-\frac{1}{N}\sum_{i=1}^{i=N}\sum_{j=1}^{j=M} y_{i,j} \log p_{i,j}, \quad (5)$$

where $N$ is the number of test set files, $M$ is the number of classes, and $y_{i,j}$ is the indicator variable whose value is 1 if test instance $i$ belongs to class $j$ or 0 otherwise. The rationale of using the logloss metric rather than accuracy is that logloss assesses better model robustness since it takes into account not only the model's classification decision but also the level of confidence with which it is made. It is well-known that even very accurate random forest classifiers may output class probabilities of poor quality [45], [46]. Consequently, when optimizing towards logloss rather than accuracy, we apply a standard technique for calibrating the probabilities output by DAEMON's random forest model [45].[7]

---

[6]This was empirically found to provide better results than using the total number of occurrences as features.

**TABLE 1.** Microsoft's dataset malware families.

| Name | # Train Samples | Type |
|---|---|---|
| Kelihos_ver3 | 2,942 | Backdoor |
| Lollipop | 2,478 | Adware |
| Ramnit | 1,541 | Worm |
| Obfuscator.ACY | 1,228 | Any kind of obfuscated malware |
| Gatak | 1,013 | Backdoor |
| Tracur | 751 | TrojanDownloader |
| Vundo | 475 | Trojan |
| Kelihos_ver1 | 398 | Backdoor |
| Simda | 42 | Backdoor |

The number of teams that participated in the competition is over 370 and the best (smallest) logloss, achieved by the winning team, is 0.00283. Although the competition was completed on April, 2015, at the time of this writing, the submission site still accepts late submissions and returns their logloss.

### B. THE DREBIN DATASET

The Drebin dataset [17], [31] is a collection of 131,611 Android applications, the majority of which are benign. Applications are in Dalvik executable format. It is widely used as a benchmark for both malware classification and detection [6], [7], [17], [19], [22]–[24], [31], [47], [48]. The majority of the applications were collected from Google Play. Drebin also includes all samples from the Android Malware Genome Project [37]. In terms of malware, Drebin contains 5,560 malicious applications from 179 malware families of widely-varying sizes. Since the majority of these families are very small (less than 10 samples), we adopt the approach taken by previous malware classification works that have used Drebin [22]–[24] and consider only families of minimum size. Table 2 presents the 24 malicious Drebin families that contain 20 or more samples, which we use in our evaluation. These 24 families collectively contain 4,783 malicious samples. Families whose names appear in **boldface** are SMS-Trojan families. In Section IX, we analyze in detail how DAEMON succeeds in distinguishing between these families.

**TABLE 2.** Drebin dataset: malware families of size 20 or more.

| Name | # Samples | Name | # Samples |
|---|---|---|---|
| **FakeInstaller** | 925 | ExploitLinuxLotoor | 70 |
| DroidKungFu | 667 | GoldDream | 69 |
| Plankton | 625 | **MobileTx** | 69 |
| **Opfake** | 613 | FakeRun | 61 |
| GinMaster | 339 | SendPay | 59 |
| BaseBridge | 330 | Gappusin | 58 |
| Iconosys | 152 | Imlog | 44 |
| **Kmin** | 147 | SMSreg | 41 |
| FakeDoc | 132 | **Yzhc** | 37 |
| Geinimi | 92 | Jifake | 29 |
| Adrd | 91 | Hamob | 28 |
| DroidDream | 81 | **Boxer** | 27 |

### C. THE CIC-InvesAndMal2019 DATASET

CIC-InvesAndMal2019 [32], [33] is a dataset of 10,854 samples, 426 of which are malicious applications found on real devices. Each of the malicious applications belongs to one of four different categories: Adware, Ransomware, Scareware, and SMS Malware, from a total of 42 different malware families. Tables 3-6 present the malicious families of the dataset in each malware category, which we use in our evaluation. In addition to the application files themselves, the dataset also contains a set of pre-computed features extracted from them. These features include dynamic features such as network traffic features extracted from PCAP files, as well as API-Calls related features, and static features such as permissions and intents. Although this dataset is rather small, DAEMON succeeds in both distinguishing between its malicious families and in malware categorization.

### VII. EXPERIMENTAL EVALUATION

In this section, we present the results of DAEMON's experimental evaluation on the three datasets. Since we have tuned

---

[7] A similar calibration technique was applied by the team that finished the competition in the 7th place.

**TABLE 3.** CIC-InvesAndMal2019 dataset: Adware families.

| Name | # Samples | Name | # Samples |
|---|---|---|---|
| Dowgin | 10 | Ewind | 10 |
| Feiwo | 15 | Gooligan | 14 |
| Kemoge | 11 | Koodous | 10 |
| Mobidash | 10 | Selfmite | 4 |
| Shuanet | 10 | Youmi | 10 |

**TABLE 4.** CIC-InvesAndMal2019 dataset: Ransomware families.

| Name | # Samples | Name | # Samples |
|---|---|---|---|
| Charger | 10 | Jisut | 10 |
| Koler | 10 | LockerPin | 10 |
| Simplocker | 10 | Pletor | 10 |
| PornDroid | 10 | RansomBO | 10 |
| Svpeng | 11 | WannaLocker | 10 |

**TABLE 5.** CIC-InvesAndMal2019 dataset: Scareware families.

| Name | # Samples | Name | # Samples |
|---|---|---|---|
| AndroidDefender | 17 | AndroidSpy.277 | 6 |
| AV | 10 | AVpass | 10 |
| FakeApp | 10 | FakeApp.AL | 11 |
| FakeAV | 10 | FakeJobOffer | 9 |
| FakeTaoBao | 9 | Penetho | 10 |
| VirusShield | 10 | | |

**TABLE 6.** CIC-InvesAndMal2019 dataset: SMS families.

| Name | # Samples | Name | # Samples |
|---|---|---|---|
| Biige | 11 | FakeInst | 10 |
| FakeMart | 10 | FakeNotify | 10 |
| Jifake | 10 | Mazarbot | 9 |
| Nandrobox | 11 | Plankton | 10 |
| SMSsniffer | 9 | Zsone | 10 |
| BeanBot | 9 | | |

DAEMON's parameters using Microsoft's dataset, we start by describing our evaluation results on this dataset.

### A. EVALUATION RESULTS ON MICROSOFT's DATASET

We described Microsoft's dataset in Section VI-A. We remind the reader that each of the training and test sets comprises approximately 11,000 files. The test set was further (randomly) partitioned by Microsoft into two subsets: the **public test set** (comprising 30% of the test set) and the **private test set** (comprising 70% of the test set). At the end of the competition, contestants were ranked in increasing order of the logloss (see Equation 5) obtained by their model on the private test set and results were made public on the **private leaderboard**.[8] In order to provide contestants with some feedback on their relative performance on test files while the contest was ongoing, a **public leaderboard** was made available to them, ranking models based on the public test set. We did not use the public leaderboard in our model generation process.

[8]https://www.kaggle.com/c/malware-classification/leaderboard

Microsoft rated contestants only based on the logloss of their classification models. However, many contestants, as well as malware classifiers that were trained using this dataset after the competition was completed (such as DAEMON), evaluated their models also (or only) by computing $k$-fold cross-validation accuracy on the training set. Therefore, we evaluated DAEMON using both logloss and (5 fold) cross-validation accuracy.

Recall that DAEMON uses two sets of features: 1-grams and $N$-grams, for $N \in \{4, 8, 16, 32\}$. In order to measure the extent to which each of these sets contributes to DAEMON's performance, we define and evaluate two variants: DAEMON-1G uses only the 1-gram features, whereas DAEMON-NG uses only the $N$-grams, for $N \in \{4, 8, 16, 32\}$. Table 7 presents the results of the 5 best-performing models on Microsoft's dataset in terms of logloss, along with the results of DAEMON's two variants. Recall that over 370 teams have participated in the competition and more than 30 additional models were evaluated on the dataset afterward, for an overall of more than 400 classification models.

**TABLE 7.** Comparison of DAEMON with top classifiers on Microsoft's dataset.

| Classifier | Rank | Logloss | CV Accuracy |
|---|---|---|---|
| Winning Team | 1 | 0.00283 | 99.83 |
| Marios & Gert Team | 2 | 0.00324 | N/A |
| DAEMON | 3 | 0.00391 | 99.72 |
| Mikhail, Dmitry and Stanislav | 4 | 0.00396 | N/A |
| Zhang et al. [13] | 5 | 0.00426 | 99.79 |
| Octo Guys Team | 6 | 0.00519 | N/A |
| Ahmadi et al. [11] | 7 | 0.0063 | 99.77 |
| DAEMON-NG | 18 | 0.00730 | 99.66 |
| DAEMON-1G | 86 | 0.02235 | 99.1 |

Focusing first on logloss, DAEMON is ranked 3rd out of all models with a logloss that exceeds that of the winning team by only approx. 0.001. We note that the vast majority of classifiers *are not platform-agnostic*. Moreover, to the best of our knowledge, none of the first 10 most highly-ranked classifiers except DAEMON are platform-agnostic. For example, the classifier of the winning team uses features such as op-code counts and segments count and that of the second-ranked team used features such as the number of lines in each PE section. Turning out attention to DAEMON's variants, we see that the model based on the $N$-grams set of features (ranked 18) is more powerful than that based on the 1-gram features (ranked 86). Nevertheless, both are required for DAEMON to perform as well as it does.

DAEMON's cross-validation accuracy is 99.72 which is very high, but slightly lower than that of the winning team's model and that of the models of [11], [13]. The team ranked 2nd did not report on cross-validation accuracy. Out of all the contestants and later works[9] that reported on classification accuracy, DAEMON's accuracy is ranked 4th. The N-grams (for $N > 1$) used by DAEMON are stronger than 1-grams also in terms of accuracy, but it is their combination that performs best.

[9]Ronen *et al.* [30] reports on more than 50 research papers published during 2015-2018 that have used the Kaggle competition's dataset.

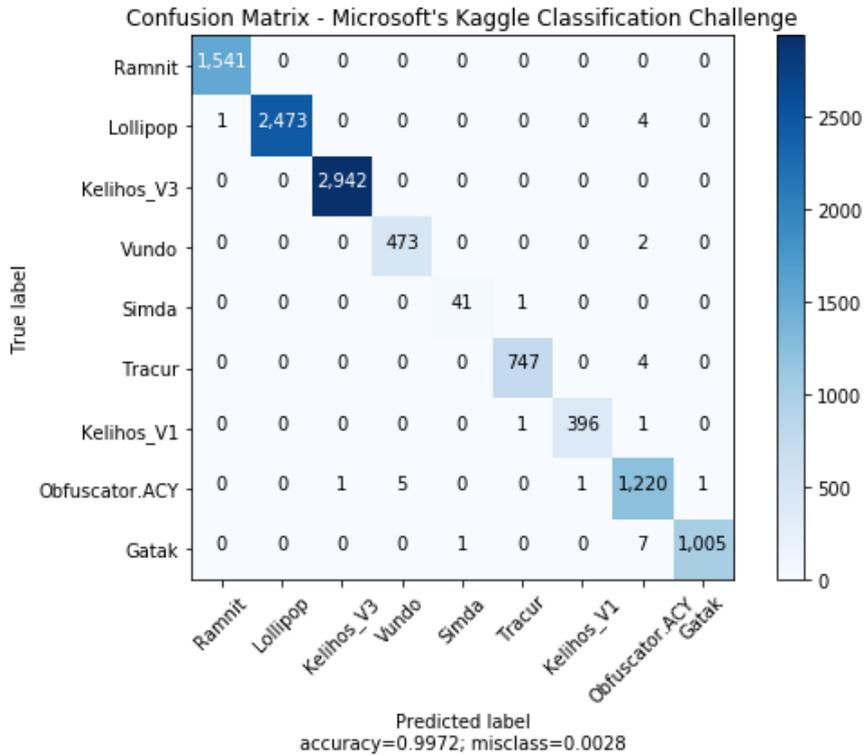

**FIGURE 2.** DAEMON's 5-Fold CV Confusion Matrix on Microsoft's dataset.

Figure 2 presents DAEMON's confusion matrix on the dataset. Although the dataset is very imbalanced, even the smallest family – Simda – is classified with high accuracy (97.6%).

We remind the reader that two files are provided for each sample in Microsoft's dataset: the sample's PE binary file and a corresponding disassembly file. The evaluation results reported above were obtained by inputting both files to the classifiers. Disassembly files are constructed from binaries based on knowledge of a platform's instruction-set and the binary's structure and semantics. Consequently, one may argue that although DAEMON does not directly use any platform-dependent features, its power as a platform-agnostic classifier would be better assessed when applied to binary files only rather than receiving also disassembly-file. Thus, we have also trained DAEMON using only binary files.

Table 8 presents the logloss and accuracy results of DAEMON and the few classifiers that received as their input the dataset's binary files only [14]–[16]. Like DAEMON, all these 3 classifiers only use platform-agnostic features. Unlike DAEMON, they all employ deep learning architectures. As can be seen, DAEMON obtains very high accuracy and very low logloss also when trained on and applied to binary files only. It also significantly outperforms

**TABLE 8.** Comparison of DAEMON with platform-agnostic classifiers using binaries only on Microsoft's dataset.

| Classifier | Logloss | CV Accuracy |
|---|---|---|
| **DAEMON** | **0.0107** | **99.56** |
| Kebede et al. [14] | N/A | 99.15 |
| Narayanan et al. [16] | 0.0774 | 98.2 |
| Le et al. [15] | N/A | 96.6 |

all other platform-agnostic classifiers in terms of both accuracy and logloss.

### B. EVALUATION RESULTS ON THE DREBIN DATASET

As we've described in Section VI-B, the Drebin dataset is very imbalanced and the majority of its 179 families are too small for classification purposes, as they contain less than 10 samples. Consequently, as done by previous works with which we compare DAEMON [22]–[24], we have conducted our evaluation by using only families that contain at least 20 samples and have randomly divided the dataset consisting of these families into a training set (consisting of 70% of the samples) and a test set (consisting of 30% of the samples).

Figure 3 presents DAEMON's confusion matrix on Drebin based on its classification results on the test set. As can be seen, even though very small families have been removed, the remaining dataset is still very imbalanced. Nevertheless, DAEMON achieves high accuracy even on small families. For instance, it achieves 100% accuracy on both the YZHC and Boxer families, that have only 7 and 8 test-set samples, respectively.

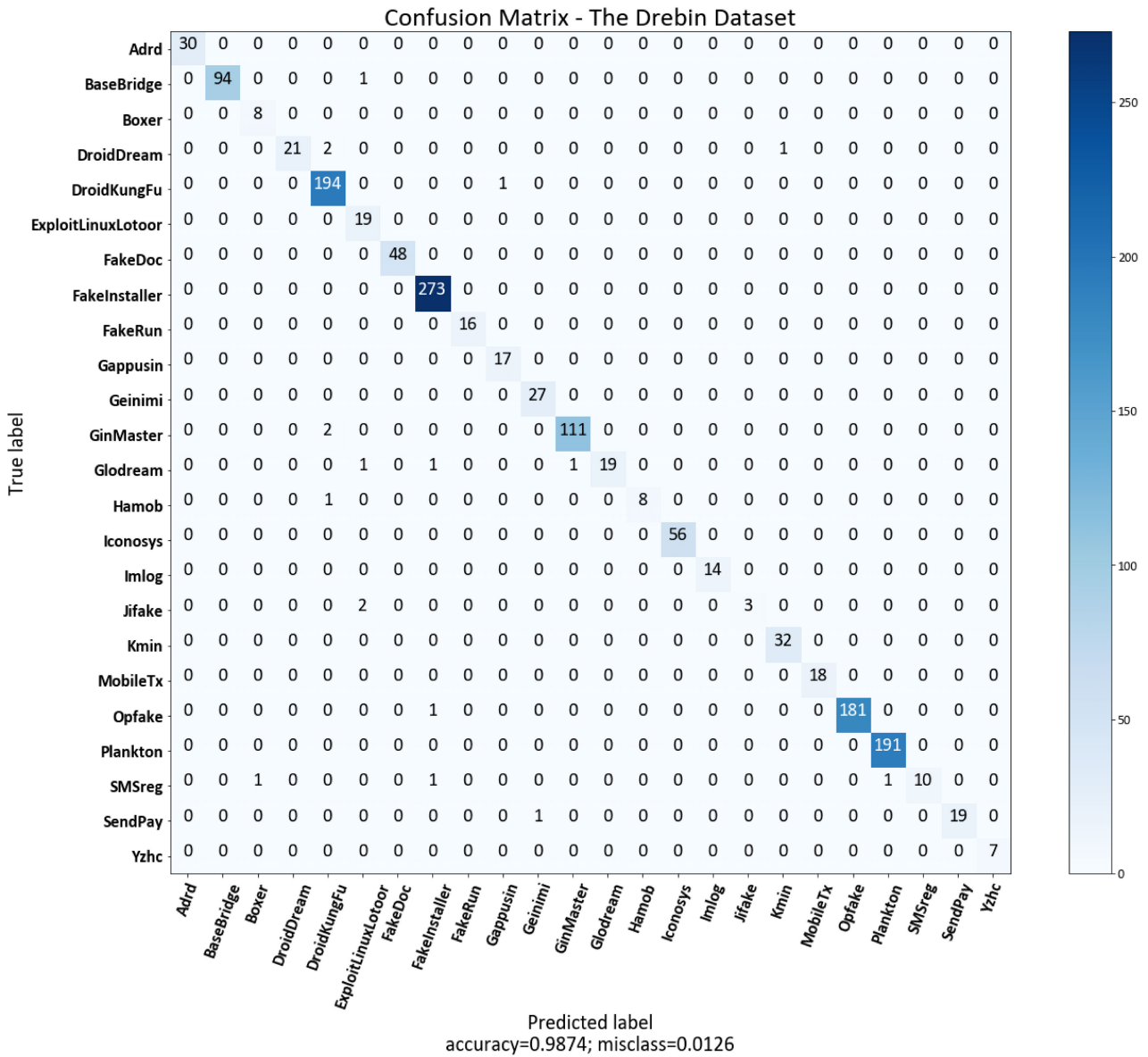

**FIGURE 3.** DAEMON's Drebin's confusion Matrix.

Table 9 compares the accuracy of DAEMON with that of previously published malware classifiers that were evaluated using DREBIN. DAEMON obtains high accuracy of 98.74% on the test set. DAEMON's accuracy is almost 15 pp. more than the 2nd best classifier for which this experiment was conducted [22]. We note that DroidSieve [7], a static-analysis based Android malware detector, achieved 98.12% accuracy in family identification on a subset of files it detected as malicious. These files belong to 108 families out of the 176 families of the Drebin dataset. Unlike DAEMON, DroidSieve uses domain-specific knowledge and platform-dependent features.

**TABLE 9.** Comparison of DAEMON with other classifiers on the Drebin dataset.

| Classifier | Accuracy |
|---|---|
| DAEMON | **98.74** |
| Dash et al. [22] | 84 |
| Massarelli et al. [24] | 82 |
| Le et al. [23] | 81.8 |

We emphasize that we have optimized DAEMON using Microsoft's dataset and that it has been applied to the Drebin dataset without any algorithmic changes, feature re-engineering, or parameter tuning. Thus, these results establish empirically that DAEMON is an effective dataset-agnostic, as well as platform-agnostic, malware classifier.

### C. EVALUATION RESULTS ON THE CIC-InvesAndMal2019 DATASET

As shown by Tables 3-6, the malicious part of the CIC-InvesAndMal2019 dataset is quite balanced in comparison with DREBIN. On the other hand, it is much smaller and contains only 426 malicious executables. Three experiments were conducted on this dataset by the researchers of the Canadian Institute for Cybersecurity (CIC) [32], [33]. They evaluated the performance of their method on the tasks of malware detection, malware category classification, and malware family classification.

Since DAEMON is a malware classifier rather than a malware detector, we compare its performance with that of the CIC method in terms of malware categorization and family classification. This is done by comparing DAEMON's results with the results reported by [32], [33], in their second and third experiments. All tests conducted by [32], [33] were based on classifiers that use popular machine learning algorithms.

In their first work [32], they suggested a dynamic malware classifier and used it in their second and third experiments, testing it with the tasks of malware categorization and family-classification, using 80 different network-flow features. In their second work [33] they added dynamic features based on API-calls as well, and also improved the performance of their malware detector.

Imtiaz *et al.* [9] presented DeepAMD, a deep learning based Android Malware Detector, which was evaluated on the CIC-InvesAndMal2019 dataset. Similarly to the CIC method, they chose to train their model on the features (both static and dynamic) provided with the dataset. They didn't use any additional features extracted from the applications. They evaluated their model on the tasks of malware categorization and family classification using similar experiments to the ones performed by CIC. We compare DAEMON with both these classifiers.

#### 1) CIC - EXPERIMENT 2 (Malware Categorization)

In this experiment dataset executables that were classified as malicious in a previous experiment are classified into their respective categories:

Adware/Scareware/Ransomware/SMS Malware. The researchers randomly split the dataset into 80% training set and 20% test set, and evaluated their performance on the test set. More recently, In [33], they tested several machine learning algorithms trained using only the dynamic features provided with the dataset. Their results show that the random forest based model outperforms the rest of the models. This model obtained a precision of 83.30 and a recall of 81.00 in this experiment. DAEMON, tested on all the malicious files in the test set, without using any of the static or dynamic features provided with the dataset, obtained a precision of 92.21 and recall of 91.74.

#### 2) DeepAMD - MALWARE CATEGORIZATION

DeepAMD was evaluated using an experiment similar to experiment 2 performed by the CIC. In this experiment, the researchers attempted to categorize the entire dataset, including benign applications. We remind the reader that the dataset contains 5,065 benign applications and 426 malicious applications. After splitting the dataset as done by the CIC, they categorized each application as one of the following: Benign/Adware/Scareware/

Ransomware/SMS Malware/Premium SMS.

In this test they used the static features provided with the dataset and achieved a precision of 92.2 and recall of 92.5.

In an additional experiment, DeepAMD was evaluated in categorization using only the files classified as malicious in their malware detection experiment. In this experiment, the researchers trained a model using only the dynamic features provided with the dataset. Therefore, *this experiment fits the exact setting of experiment* 2 *performed by the CIC*. DeepAMD achieved a precision of 82.2 and a recall of 80.3 in this experiment. Hence, DAEMON managed to outperform DeepAMD in this task, even without using dynamic features.

#### 3) CIC - EXPERIMENT 3 (Malware FAMILY Classification)

In this experiment, dataset executables were classified into their respective families. As in the previous experiments, the dataset was randomly split into 80% training set and 20% test set. In [33], they trained a random forest model using dynamic features provided with the dataset and evaluated it on the test set. Their model obtained a precision of 59.70 and recall of 61.20..[10] In this exact task, DAEMON obtained a precision of 83.56 and a recall of 77.64.

#### 4) DeepAMD - MALWARE FAMILY CLASSIFICATION

DeepAMD conducted several experiments regarding malware family classification. One of these experiments is *essentially the same as experiment* 3 *performed by the CIC*. In this experiment, they achieved a precision of 65 and a recall of 59. Hence, DAEMON significantly outperformed DeepAMD as well. We remind the reader that in accordance with the previous experiments, DAEMON uses only the applications themselves as inputs. Thus, unlike DeepAMD, it doesn't use any feature (static or dynamic) provided with the CIC-InvesAndMal2019 dataset.

### VIII. TIME COMPLEXITY

An important factor in assessing a practical malware classifier is the extent to which it can scale to large sample collections in terms of time-complexity. We now report on the time it takes DAEMON to learn a model and to classify new samples on the three datasets. All of our experiments were conducted on a 2.00 GHz 24-core Xeon E5-2620 server, with 256GB RAM, running the 64-bit Ubuntu 14.04 operating system.

---

[10]Note that classification accuracy is drastically lower than that of Experiment 1, because classifying into families is harder than classifying into categories, each consisting of multiple families

### A. MICROSOFT's DATASET

Tables 10 and 11 respectively present DAEMON's model generation times on binaries only and on both binary/disassembly files. The tables present the time it takes to perform each of DAEMON's model generation stages based on all of the dataset's 10,868 training samples. Focusing first on the model generated from binaries only, we see that the most time-consuming stage is that of extracting family-representative $N$-grams (stage 2). This stage takes slightly over 7 hours to complete. Total model generation time is approximately 11 hours. The time required for generating a model using both the binary and the disassembly files is naturally longer. In this case as well, stage 2 is the most time consuming and takes slightly less than 11 hours. The total model generation time is approximately 16 hours.

**TABLE 10.** Model generation times: Microsoft's binaries.

| Stage | Time (minutes) |
|---|---|
| Entropy Threshold Computation | 3.3 |
| Family Representative N-grams Extraction | **438** |
| Pairwise-Separating Features Selection | 6.3 |
| Feature Vectors Computation | 105 |
| Random Forest Model Generation | 8 |

**TABLE 11.** Model generation times: Microsoft's binaries + disassembly files.

| Stage | Time (Minutes) |
|---|---|
| Entropy Threshold Computation | 5.8 |
| Family Representative N-grams Extraction | **653** |
| Pairwise-Separating Features Selection | 17 |
| Feature Vectors Computation | 185 |
| Random Forest Model Generation | 9 |

As for DAEMON's classification times, we have measured them when using only binary files and when using both binary and disassembly files. The overall time it took DAEMON to classify all of the dataset's 10,873 test files *on a single core* was approximately 176 minutes, translating to a classification rate of approximately 62 files per minute.

### B. THE DREBIN DATASET

The Drebin dataset is considerably smaller than Microsoft's dataset in terms of both the number of dataset files and their average size. Consequently, the time it takes to train a DAEMON model or to classify a new sample is much smaller as well. The time it took to perform each of DAEMON's model generation stages is presented in Table 12. The total time it took to build and train the model, in this case, was slightly less than 2 hours. The classification rate *on a single core* was approximately 240 files per minute.

### C. THE CIC-InvesAndMal2019 DATASET

The CIC-InvesAndMal2019 is rather small. Nevertheless, some of the families in it contain many candidate strings and thus were harder to process in comparison with most of the Drebin dataset families. As can be seen in Table 13, the most time-consuming part for this dataset was the computation of the pairwise-separating features, since there are 42 families in the dataset. This required mining features for many more family-pairs in comparison with our experiments with both the Drebin dataset and Microsoft's dataset. Category classification times are shown by Table 14. The classification rate on this dataset *on a single core* was approximately 84 files per minute.

**TABLE 12.** Model generation times: DREBIN.

| Stage | Time (Minutes) |
|---|---|
| Entropy Threshold Computation | 0.3 |
| Family Representative N-grams Extraction | **59** |
| Pairwise-Separating Features Selection | 17.2 |
| Feature Vectors Computation | 30.9 |
| Random Forest Model Generation | 7 |

**TABLE 13.** Model generation times: CIC-InvesAndMal2019 family classification.

| Stage | Time (Minutes) |
|---|---|
| Entropy Threshold Computation | 0.2 |
| Family Representative N-grams Extraction | 183 |
| Pairwise-Separating Features Selection | **188** |
| Feature Vectors Computation | 14.5 |
| Random Forest Model Generation | 3 |

**TABLE 14.** Model generation times: CIC-InvesAndMal2019 malware categorization.

| Stage | Time (Minutes) |
|---|---|
| Entropy Threshold Computation | 0.3 |
| Family Representative N-grams Extraction | **180** |
| Pairwise-Separating Features Selection | 36 |
| Feature Vectors Computation (single core) | 16.8 |
| Random Forest Model Generation (single core) | 3 |

Model generation times for all 3 datasets range between 2-16 hours, which allows even daily re-training. Detection times *on a single core* range between 62 files per minute (on Microsoft' dataset) and 240 files per minute (on the DREBIN dataset). This rate scales linearly with the number of available cores, since every test sample can be classified independently of other samples. Thus, on our 24-core server, classification rates for Microsoft's dataset, the DREBIN and the CIC-InvesAndMal2019 datasets are, respectively, 1,488, 5,760, and 2,016 files per minute.

### IX. LEVERAGING DAEMON's FEATURES FOR EXPLAINABILITY

The vast majority of DAEMON's features are relatively long $N$-grams extracted from a malicious sample's contents. As such, they are often much more useful than statistical features for gaining insights into the behavior of malware families. Specifically, they can be used for gaining insights into which vulnerabilities are exploited by the malware and how it attempts to avoid detection. These features are often strings that reveal which dynamic-link libraries and API-calls are used by a malware. Other common types of features, seen in Android malware datasets, include permissions that are requested by malicious applications and the URLs or IPs they communicate with. In other cases, $N$-grams represent binary code snippets that serve as effective family signatures.

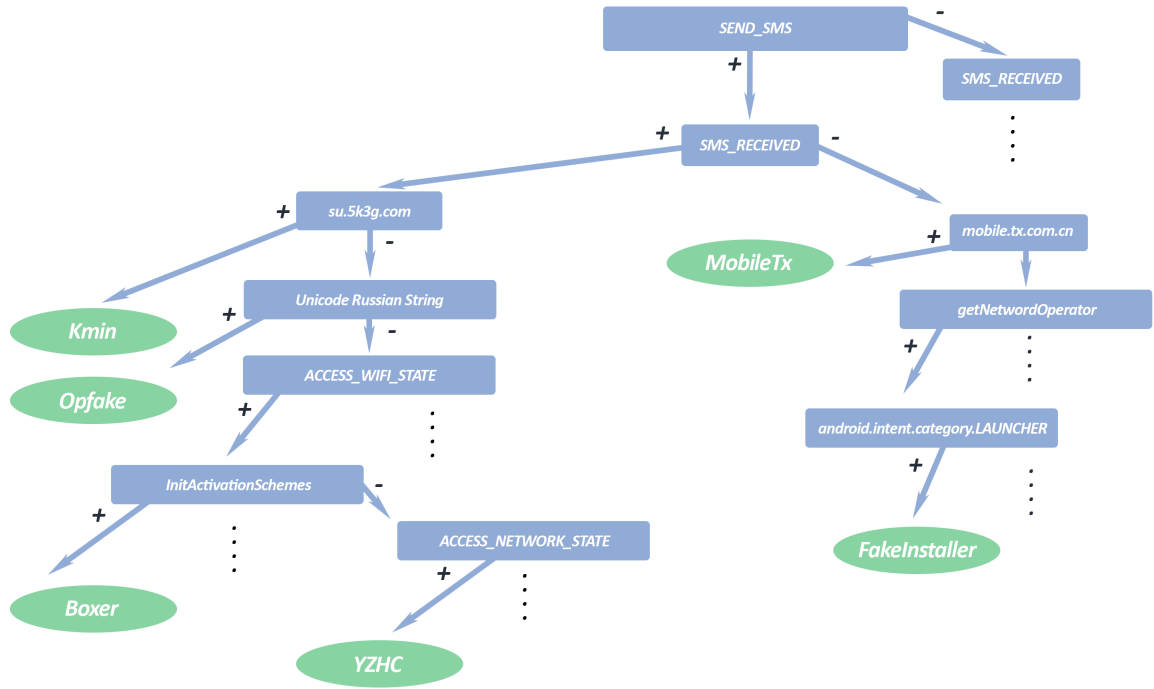

**FIGURE 4.** Features identifying SMS-Trojan variants (simplified).

These $N$-grams are extracted by DAEMON in a platform-agnostic manner, without any knowledge of the executable's format. Moreover, as it turns out, they can also be extracted from malware families that are encrypted and/or packed. Let us also recall that a by-product of DAEMON's feature mining process is that each $N$-gram is tagged by the family-pairs for which it obtained relatively high information gain. This makes it remarkably easier to identify the key distinguishing features of each family. In what follows, we demonstrate DAEMON's explainability via examples from the three datasets on which we evaluated it.

### A. THE DREBIN DATASET
#### 1) SMS-TROJANS
We've evaluated DAEMON using Drebin's 24 malware families that contain at least 20 samples. Six of these families are different types of SMS-Trojans (see boldface names in Table 2): `FakeInstaller`, `Opfake`, `Kmin`, `MobileTx`, `Yzhc`, and `Boxer`. SMS-trojans use the SMS services of an Android device for sending and/or intercepting SMS messages for malicious purposes. They differ, however, in their goals and attack tactics. In our first example of DAEMON's explainability, we describe the key features it uses for effectively telling SMS-Trojans apart from other families and for differentiating between different types of SMS-Trojans.

Figure 4 depicts a decision-tree-like structure (which is a simplification of the actual model) showing some (but not all) of the features we describe below that are used by DAEMON for classifying SMS-Trojans. A '+' sign indicates that the corresponding $N$-gram was found in the classified sample and a '−' sign indicates it is absent from it.

`Kmin`
In addition to sending SMS messages to premium-rate numbers, variants of this family also download and install other applications onto the victim's device. Moreover, as stated in Microsoft's report on this family [49], its variants send the following data to their C&C server: "Device ID", "Subscriber ID" and "Current Time". DAEMON extracted all these strings as top-most separating features between `Kmin` and other SMS-Trojan families. The names of the functions used in order to obtain this data, such as "getSubscriberId" and "getDeviceId", were extracted as well. Collectively, these features identify `Kmin` variants by revealing the type of data they aim to exfiltrate. Another strong feature used by DAEMON for identifying `Kmin` variants is the string "http://su.5k3g.com/", which is a URL of a remote server with which only variants of this family communicate.

Another example of a behavioral feature that can be used for identifying Kmin variants is the string "telephony.sms_SMS_RECEIVED". This feature indicates that the application requests to be notified when an SMS is being received. This mechanism is used by family variants as an evasion mechanism, to block messages from the mobile operator regarding phone charges that are being made. This way smartphone users are being kept in the dark

w.r.t. their charges [37]. Another indicative feature is the "vnd.wap.mms-message" feature, that is also a top-most separating feature between `Kmin` and all other SMS-Trojan families. It indicates that Kmin variants send MMS messages (in addition to SMSs) to premium-rate numbers. The combination of all these features (as well as others) allows DAEMON to classify `Kmin` variants with high precision.

MobileTX

The MobileTx family also steals data from the compromised device in addition to sending SMS messages to premium-rate numbers. Stolen data is sent to an account hosted by the following remote server URL, extracted by DAEMON as a top-most separating feature between `MobileTX` and all other SMS-Trojans: "mobile.tx.com.cn" [50]. Stolen data includes the smartphone's IMEI and phone-number and DAEMON designates strings identifying these types of data as top-most features separating the `MobileTX` family from other SMS-Trojan families as well. Another helpful feature extracted by DAEMON is the string "com/tx/bean/TxMenu", which contains the name of a package used exclusively by variants of the `MobileTX` family, most probably for communication with its remote server.

YZHC

YZHC variants send premium-rate SMS messages and block all incoming messages that inform the user about it. Another malicious behaviour of this family is the exfiltration of private information from the device. One feature that was designated by DAEMON as a top-most separating feature between `YZHC` and all other SMS-Trojan families, except `Kmin`, is "PackageInstaller". A top-most separating feature between `YZHC` and `Kmin` is ACCESS_NETWORK_STATE. This feature is the name of a permission used by `YZHC` variants for obtaining data regarding the communication network, which is used by them (in order to decide when to send premium SMS messages) [51] but is not used by `Kmin` variants.

FakeInstaller

Unlike most other SMS Drebin Trojan families, variants of `FakeInstaller` collect data regarding the cellular operator [52]. DAEMON uses the "getNetworkOperator" N-gram as a top-most separating feature for this family, which is indicative of this behavior.

Boxer

Boxer is a family of malware that pretends to be an installer or application downloader, but in reality, sends premium-rate SMS messages without the user's acknowledgement. A distinguishing feature of `Boxer` variants is that they use Android Cloud to Device Messaging (C2DM) services for communicating with a cloud-based C&C server. Indeed, DAEMON uses the "C2DM" and `C2DM_INTENT` features (not shown in Figure 4) for identifying them. In addition, variants of this family are able to target multiple countries.

In order to do so, they call (among other functions) the "InitActivationSchemes" function, an N-gram feature used by DAEMON. This function is used to match the Mobile Country Code they previously read to a proper identifier so that SMS messages can be correctly sent from each such country.

Opfake

The Opfake family mostly targets Russian smartphones [53]. After being downloaded to the device, variants display a service agreement message to the user in Russian that describes the usage of paid SMS messages. Consequently, one of the top-most separating features between this family and all other SMS-Trojans is a short Unicode Russian text taken from the bytecode that is part of this agreement's text.

Plankton

Moving on from SMS-Trojans, we next discuss the `Plankton` malware family, discovered in June 2011 [54]. Plankton variants download their malicious payload from a remote server, unlike most other Drebin families.

Since family variants communicate with their remote server using "HTTP POST" messages, DAEMON is able to extract top-most separating HTTP-related features. An Example of such a feature is "apache/http/post".

Moreover, DAEMON also extracted features identifying which data is collected by family variants, such as "getIMEI" for exfiltrating the device IMEI, "getDisplayMetrics" for discovering the user's display resolution, etc.

Some of the family variants require access to the Internet and to the WiFi state as well. This is used for accessing the list of contacts, the history of calls and browser bookmarks that are then communicated to the remote server. DAEMON extracts features revealing this behavior, such as "WifiManager", "WiFi", and many more. The combination of these features allows DAEMON to accurately classify `Plankton` variants.

GoldDream

`GoldDream` is a family of Android Trojans that monitor an infected device and collect sensitive data over time. After the malware has collected sufficient data, it sends it to a C&C server, whose hostname has also been extracted by DAEMON: "lebar.gicp.net". This was designated by DAEMON as a top-most separating feature between `GoldDream` and all other families.

### B. MICROSOFT'S MALWARE CLASSIFICATION CHALLENGE DATASET

We remind the reader that this dataset consists of 9 malware families of different types (Worms, Adwares, Backdoors, Trojans, etc). We note that due to the fact that Microsoft has removed the headers from the PE executables in the dataset, it is difficult to decipher the meaning of many N-grams extracted from packed/encrypted content. In the following, we examine some of the top-most separating features extracted by DAEMON for a large family, W32.Ramnit, and explain how these features shed light on the behavior of this family.

Ramnit (W32.Ramnit)

Ramnit is a worm that spreads through removable drives on Windows x86 systems, infecting EXE and DLL files. The primary goal of its polymorphic variants is stealing information such as cookies, in order to hijack online sessions with banking and social media websites. Moreover, family variants open a connection to a C&C server in order to receive commands instructing them to perform various operations. Examples of such operations include capturing screenshots, uploading stolen cookies, deleting root registry keys, preventing the PC from starting up, etc. Although most of their content is encrypted, DAEMON manages to extract from Ramnit files high-quality separating features. For instance, it appears that there are some encrypted/packed parts of the malware which can be found *in all of its dataset variants*, such as the byte-sequences "C9C35651538D9920" and "C71083C11039D175C9". Combining these features creates a perfect family signature, identifying its variants with perfect accuracy.

Another top-most separating feature is "GetCurrentProcess", an API call used by Ramnit to walk the stack and suspend threads of "rapportgp.dll", a lightweight security software designed to protect confidential information from being stolen by malware. DAEMON also extracts the features: "LoadLibraryA" and "GetProcAddress" from PE files. Both features are a maliciousness indicator since malware often uses these API-calls in order to load DLL files whose names do not appear in their PE header, often containing the malicious payload. This is rarely done by benign applications. As stated in Symantec's detailed report on Ramnit [55], its variants use these API-calls as part of their client infection process.

### C. CIC-InvesAndMal2019 DATASET

Similarly to Drebin but unlike Microsoft's dataset, applications' headers were not removed in the CIC-InvesAndMal2019 dataset, which facilitates explainability. In what follows, we shortly analyze the FakeAV family (which belongs to the Scareware category) and explain how the features extracted by DAEMON from its files shed light on its malicious payload.

FakeAV is a malware family that spreads under the disguise of popular Android applications. After installation, the malware alerts victims regarding security threats that do not exist on their Android device and recommends that they visit a website where they will be asked to pay for cleaning these threats. Additionally, it can be used by a C&C server to perform many actions: send messages, make calls, open a URL, install applications, etc [56]. Moreover, upon receiving a command, the malware sends information about contacts, call history, current location, and account information details.

Many of DAEMON's top separating features for this family expose this behavior. For instance, the features "Android/telephony/CellLocation" and

"Android/telephony/gsm/GsmCellLocation" expose the malware's attempts to obtain the current location of the device. In addition, the features

"action.NEW_OUTGOING_CALL" and

"Android.intent.extra.PHONE_NUMBER" further indicate the application's intent to obtain the phone number and monitor when outgoing calls are being made. DAEMON also selected "telephony/SmsManager" as a top separating feature for FakeApp, shedding light on the real intentions of FakeAV variants.

## X. DISCUSSION

Being dataset-agnostic is a significant advantage of a malware classifier, as it allows successfully applying it out-of-the-box to new malware collections. Such collections may result from the availability of new data and/or the appearance of new malware families. In addition, new malware collections may also result from the emergence of new computing platforms or new executable formats, although this event is rare. While platform-agnostic malware classifier can be applied to such collections out-of-the-box, platform-dependent classifiers cannot. Classifiers that rely heavily on platform-dependent features will likely require extensive feature re-engineering, optimization, and tuning.

DAEMON was trained and optimized using a dataset consisting of Windows executables and was then successfully applied to two datasets of Android applications. This process was done without any algorithmic changes, feature re-engineering, or parameter tuning. Although it establishes that DAEMON is platform-agnostic, there is obviously no guarantee that DAEMON will provide top-notch performance for all existing or future such datasets. Consequently, in future work, we plan to try to obtain additional malware datasets (possibly also of additional platforms) in order to evaluate DAEMON's performance and ascertain that it can be successfully applied to them with no (or at least with minimum) changes.

DAEMON can be deployed by malware analysts of large organizations or anti-malware vendors in order to quickly classify new malicious variants, assess the risks they pose, and determine whether further manual analysis is required.

Recent work on malware detectors focuses on their sustainability [57]–[64]. The first provably sustainable malware detector was introduced by Mariconti *et al.* [63], [64], although the term "detector sustainability" was introduced in a later work by Jordaney *et al.* [57]. The work defines a malware detector as *sustainable* if, once trained on a dataset, it can continue to effectively detect new malware without retraining for an extended period of time. Avoiding retraining, or at least reducing the frequency in which the model has to be re-trained, is advantageous, because retraining requires new training instances that typically have to be labeled manually.

The key idea of the sustainable Android malware detector presented by Mariconti et al [63], [64], MaMaDroid, is to capture an Android application's behavior using the sequence of *abstracted* API calls that it performs. API calls are abstracted to their Java class name, Java package name or source, and these sequences are modeled by Markov chains. Their evaluation shows that MaMaDroid is more robust to changes in Android malware samples and APIs that occur over an extended period of time[11] in comparison with previous state-of-the-art Android malware detectors.

Later works presented sustainable detectors whose classification quality degrades more slowly in comparison with MaMaDroid, using several techniques. Examples of such techniques are selection of more stable statistical features [58], [59], using online learning and pseudo labeling [60], and capturing semantic similarity of API calls by automatically extracting information from official API documentation [62]. While the majority of these works investigate malware detectors (binary malware classifiers), experiments done by Zhang *et al.* [62] show that their framework is also able to improve the sustainability of malware family classifiers.

Cai [19] defines sustainability in terms of two metrics: *Reusability*, that measures the extent to which a classifier is able to adapt to changes in the instances population *with* retraining, and *stability* that measures the extent to which the accuracy of a classification model that is not re-trained decays over time. DAEMON achieves excellent classification results when applied to several unrelated datasets, i.e., it is dataset-agnostic. Reusability requires that a classifier maintain fine performance when re-trained on datasets that evolve from one another. Hence, it is likely that DAEMON will fair well in terms of reusability. Since we did not evaluate the performance of DAEMON models over time without retraining, we cannot assess the extent to which it provides stability. Empirically evaluating DAEMON's stability and optimizing it is left for future work.

*Limitations and Future Work:*

Our empirical evaluation establishes that DAEMON provides high-quality classification when trained on instances from a set of malware families and used for classifying new instances from the *same* set of families. Nevertheless, even under this assumption, the evolution of malware over time may cause concept drifts and thus reduce DAEMON's predictions accuracy. This can be prevented by frequent retraining of DAEMON, but frequent retraining requires many labeled instances. This limitation of DAEMON can be mitigated by using existing frameworks that identify aging classification models [57], [65].

Jordaney *et al.* [57] presented Transcend, a statistical approach for identifying the aging of classification models during deployment. Their experimental evaluation shows that Transcend effectively identifies concept drifts in both binary and multi-class classifiers. Barbero *et al.* [65] proposes novel measures for detecting concept drift that can significantly reduce the computational cost of Transcend. Transcend is agnostic to the classification algorithm to which it is applied. In future work, we plan to investigate the extent to which Transcend can identify the concept drift of DAEMON models over time so that redundant retraining can be avoided.

Degradation in the performance of malware classifiers can stem not only from the evolution of previously known malware families, but also from the emergence of new malware families, possibly representing zero-day attacks. Integrating Transcend into DAEMON's detection pipeline may also facilitate the detection of new families, since *drifting instances* (that is, instances that are identified by Transcend as likely to be erroneously classified by the model) would typically be sent to manual analysis.

As previously mentioned, sustainability of a classification model is a function of the model's re-usability and stability. In future work, we plan to evaluate DAEMON's stability by conducting a longitudinal performance study. We also plan to investigate how DAEMON model aging can be slowed down. Observing that existing techniques for increasing sustainability [58]–[60], [62]–[64] rely on platform-dependent features, whereas DAEMON uses only platform-agnostic features, the latter research question seems challenging.

As we have discussed in Section VIII, model generation times take several hours but are sufficiently short to allow daily re-training on a commodity server for all the 3 datasets we experimented with. Classification rates on our 24-core machine ranged between 1,488-5,760 samples per minute. In deployment scenarios that require larger throughput, more cores may be used for classification. In future work, we plan to investigate ways of increasing DAEMON's classification rate. One possible way of doing so is the following: DAEMON employs the Aho-Corasick string-searching algorithm [43], which finds *all* the occurrences of each of the N-gram features within a classified sample. However, DAEMON models only require finding whether an N-gram appears in the sample or not. Optimizing the algorithm so that it only meets this weaker requirement may result in increased throughput.

An additional avenue for future work is to evaluate DAEMON's ability to classify families of related *benign* executables, such as drivers for different types of devices, multiple versions of the same software, etc. It would also be interesting to evaluate its performance on non-executable files, such as different families of mutually-related documents. Although we believe that DAEMON's feature mining is sufficiently generic to succeed also in this latter case, this may require tuning of DAEMON's parameters, such as the set of N-gram lengths.

Finally, a natural avenue for future work is to use DAEMON's effective and efficient feature mining capabilities as the basis for a novel explainable static-analysis based *malware detector* (that is, a binary classifier that decides whether a file is malicious or benign).

---

[11]Their dataset consists of malware collected over a period of 6 years.

## XI. CONCLUSION

We presented DAEMON, the first provably effective and explainable platform-agnostic (as well as dataset-agnostic) malware classifier. We evaluated it on three datasets consisting of families of malicious executables targeted to two different computing platforms: The Drebin Dataset and CIC-InvesAndMal2019, consisting of Android applications, and Microsoft's Kaggle Classification Challenge dataset, consisting of PE x86 executables. DAEMON obtained an excellent classification accuracy of 99.72% in a 5-fold cross validation applied to Microsoft's training set and came out 3rd in terms of logloss out of more than 370 different classifiers evaluated using this dataset. We then applied DAEMON, without any changes, to the Drebin dataset, where it obtained an accuracy of 98.74%, significantly outperforming all previously published malware classifiers that were evaluated on it. As for the CIC-InvesAndMal2019 dataset, DAEMON improved greatly over prior classifiers evaluated on this dataset in terms of accuracy, precision, and recall.

Furthermore, by analyzing DAEMON's classification results and selected features, one can gain powerful insights regarding the behavior of different malware families and what differentiates a malicious family from other families.

As we have shown in our analysis of DAEMON's features for different SMS-Trojan families (in Section IX), DAEMON is able to accurately classify variants even for families whose payloads are very similar. DAEMON's code is publicly available in the project's GitHub.

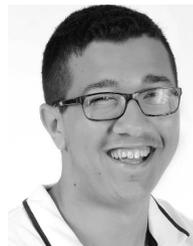

**RON KORINE** was born in Los Angeles, CA, USA. He received the B.Sc. degree in software engineering and the M.Sc. degree in computer science concurrently from Ben-Gurion University of the Negev, in 2020.

His research interest includes machine learning techniques for solving cyber security problems such as malware classification.

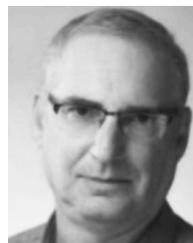

**DANNY HENDLER** (Member, IEEE) received the B.Sc. degree in mathematics and computer science and the M.Sc. degree in computer science from Tel Aviv University, in 1986 and 1993, respectively. After working in the high tech industry for 18 years in both technical and managerial positions, he returned to Academy in 2001, to pursue his Ph.D. studies. He then served as a Postdoctoral Fellow for the University of Toronto and the Technion–Israel Institute of Technology, from 2004 to 2006. He joined the Computer Science Department, Ben-Gurion University of the Negev, in 2006, where he is currently a Full Professor and the Department Chair. Since 2014, he has been serving as the Vice Head for the Ben-Gurion University of the Negev's Cyber Security Research Center. His research interests include cyber security and concurrent computing.

○ ○ ○